\documentclass[prb,nofootinbib,twocolumn,superscriptaddress]{revtex4} 

\usepackage{graphicx}
\usepackage{dcolumn}
\usepackage{bm}
\usepackage{threeparttable}
\usepackage{times}
\usepackage{mathptmx}
\usepackage{lscape}
\usepackage{natbib}
\usepackage{amsmath}
\usepackage{amssymb}
\usepackage{braket}
\usepackage{comment}
\usepackage{color}

\def\degree{\kern-.2em\r{}\kern-.3em}

\begin{document}

\title{ Structure Fluctuation Effects on Canonical-Nonlinear Thermodynamics  }

\author{Koretaka Yuge}
\affiliation{
Department of Materials Science and Engineering,  Kyoto University, Sakyo, Kyoto 606-8501, Japan\\
}%

\begin{abstract}
{  When we consider classical discrete systems under constant composition, their stable configuration in thermodynamic equilibrium can be typically obtained through the well-known canonica average $\phi$. In configurational thermodynamics, $\phi$   as a map from many-body interatomic interaction to equilibrium configuration generally exhibits complicated nonlinearity, strongly depending on their underlying lattice. The connection between nonlinearity in $\phi$ (canonical nonlinearity) and the lattice has recently been amply investigated in terms of configurational geometry, leading to establishing its stochastic-thermodynamic treatment. The present work provides natural extention of the proposed treatment, explicitly including the effect of spatial fluctuation of the equilibrium configuration on thermodynamic property of the nonlinearity. We find that the fluctuation affects the upper-bound for the averaged nonlinearity disparity in multiple configurations, as an explicit and additional contribution from stochastic mutual information between focused coordination and its fluctuation, and an implicit contribution from changes in covariance matrix for density of states due to the fluctuation. 

}
\end{abstract}


\maketitle

\section{Introduction}
For classical discrete systems under constant composition (e.g., substitutional alloys on lattice), their configuration (with prepared coordination $q=\left( q_{1},\cdots, q_{f} \right)$) in thermodynamic equilibrium can be typically obtained through the canonical average $\Braket{\quad}_{Z}$:
\begin{eqnarray}
\label{eq:can}
\Braket{q_{k}}_{Z} = Z^{-1} \sum_{d} q_{k}^{\left( d \right)} \exp\left( -\beta U^{\left( d \right)} \right),
\end{eqnarray}
where $Z$ denotes partition function, $\beta$ inverse temperature, $U$ potential energy, and summation is performed over all possible configuration $d$. When we employ generalized Ising model\cite{ce} (GIM) for the coordination, $U^{\left( d \right)}$ can be exacltly expressed as its orthonormal basis:
\begin{eqnarray}
\label{eq:gim}
U^{\left( d \right)} = \sum_{a} \Braket{U|q_{a}} q_{a}^{\left( d \right)},
\end{eqnarray}
where $\Braket{\quad|\quad}$ denotes inner product for the GIM. With these descriptions, canonical average reads as a map $\phi$ of
\begin{eqnarray}
\phi : \mathbf{Q} \mapsto \mathbf{U},
\end{eqnarray}
where $\mathbf{Q}=\left( \Braket{q_{1}}_{Z},\cdots, \Bra{q_{f}}_{Z} \right)$ and $\mathbf{U}=\left( \Braket{U|q_{1}},\cdots, \Braket{U|q_{f}} \right)$. Generally, for substitutional alloys, $\phi$ exhibits complicated nonlinearity strongly depending on their underlying lattice, i.e., its configurational geometry. 

Due to the complicated nonlinearity, various theoretical approaches have been proposed to accurately predict alloy equilibrium properties, including Metropolis algorism, entropic sampling and Wang-Landau method for effectively explore the configuration (or other appropriate) spaces, and numerical approaches to estimate the set of $\mathbf{U}$ from first-principles including genetic algorism, cross-validation and machine learning also have been amply proposed.\cite{mc1,mc2,mc3,mc4,cm1,cm2,cm3,cm4,cm5,cm6} However, these approaches do not sufficiently address how the nonlinearity is governed by its configrational geometry. 

Very recently, we provide significant progress on this issue, by introducing stochastic thermodynamic treatment of the nonlinearity:\cite{noltherm} The treatment performs transformation of system transition driven by nonlinearity into that by heat transfer from thermal bath, enabling (i) unifying the description of local and non-local contribution to the nonliearity defined on different spaces, and (ii) formulation of the nonlinearity across multiple configurations through transformed thermodynamic functions, directly relating to the lattice geometry: Especially, the latter type of formulation has not been achieved by the previous approaches. 
For instance, we derive the following inequality:\cite{noltherm}
\begin{eqnarray}
\label{eq:nth0}
\Braket{\Delta {D}_{\textrm{NOL}}}_{P_{+}^{0}} \le \ln\Braket{e^{-\sigma_{\textrm{G}}^{0}} }_{P_{+}^{0}}.
\end{eqnarray}
The equation indicates that l.h.s. of averaged nonlinearity disparity between partially ordered and other (i.e. random and ground-state ordered) configurations is bounded from above by the sum of (i) information of entropy production for artificially-constructed linear system $\sigma_{\textrm{G}}^{0}$, which can be fully estimated from information about covariance matrix of the configurational density of states (CDOS) and stochastic eigen nonlinearity (second term, discussed later). Other properties for the nonlinearity thermodynamics has also been investigated, and we find that Gibbs states act as appropriate bounds for averaged nonlinearity at thermodynamic equilibrium. 

Although the proposed thermodynamic treatment provides multiple novel insight into the nonlinearity in terms of the configurational geometry, the effect of spacial fluctuation (SF) in configuration has not been included so far. Recent progress in theoretical as well as experimental techniques enables to gradually including the effect of SF, which can significantly changes alloy configurational properties. 
The present work thus tackle this issue, explicitly including the effects of SF on nonlinearity thermodynamics, which eventually modifies bounds for averaged nonlinearity disparity of Eq.~\eqref{eq:nth0}, through additional contribution from mutual information about fluctuation and changes in covariance matrix for CDOS due to introducing the fluctuation. The details are shown below.

\section{Concepts and Derivation}
\subsection*{Nonlinearity Measure and its Thermodynamics}
Before including the effect of SF, we first briefly explain the basic concept of nonlinearity measure in terms of the configurational geometry. 
It has been shown that when CDOS takes multidimensional Gaussian distribution with the same covariance matrix $\Gamma$ of practical CDOS, $\phi$ becomes globally linear map of $\phi=-\beta\Gamma$.\cite{ig} With this consideration, local nonlinearity at given configuration $q_{J}=\left(q_{J1},\cdots, q_{Jf}\right)$ is introduced as the following vector on configuration space:\cite{asdf}
\begin{eqnarray}
\label{eq:asdf}
H\left(q\right) = \left\{\phi\circ \left(-\beta\Gamma\right)^{-1}\right\}\cdot q_{J} - q_{J},
\end{eqnarray}
where $\circ$ denotes composite map. Eq.~\eqref{eq:asdf} can be intepreted as the system evolution from configuration $q_{J}$ to $q_{J} + H\left(q\right)$ driven by the nonlinearity at $q$, $H\left(q\right)$. 

The natural extention of the above vector for nonlinearity is achieved through Kullback-Leibler (KL) divergence of\cite{ig} 
\begin{eqnarray}
D_{\textrm{NOL}}^{J} = D_{\textrm{KL}}\left(P_{J}:P^{\textrm{G}}_{J}\right),
\end{eqnarray}
where 
\begin{eqnarray}
\label{eq:cdoss}
P _{J}\left(q\right)&=& z_{J}^{-1}\cdot g\left( q \right)\exp\left[ -\beta\left( {q}\cdot {V}_{J} \right) \right]   \nonumber \\
P^{\textrm{G}}_{J} \left( {q} \right)&=& z_{J}^{\textrm{G}}\cdot g^{\textrm{G}}\left( {q} \right)\exp\left[ -\beta\left( {q}\cdot {V}_{J} \right) \right]  
\end{eqnarray}
respectively denotes canonical distribution for measuring the nonlinearity of practical and linear systems. 
$g\left( {q} \right)$ represents the CDOS of practical system with covariance matrix $\Gamma$, and $g^{\textrm{G}}\left( {q} \right)$  the CDOS of synthetically linear system, given by discretized multidimensional Gaussian with the same $\Gamma$, and we define
\begin{eqnarray}
\label{eq:ZV}
z_{J} &=&  \sum_{{q}}g\left({q} \right)\exp\left[ -\beta\left( {q}\cdot {V}_{J} \right) \right] \nonumber \\
V_{J} &=& \left( -\beta\cdot\Gamma \right)^{-1}\cdot {q}_{J}.
\end{eqnarray}
From the above equations, we can see that $V_{J}$ acts as the \textit{artificial} many-body interaction for canonical distribution to measure the nonlinearity. 
Hereafter, the superscript or subscript $\textrm{G}$ is always employed for functions of the linear system, as defined for $P^{\textrm{G}}$ and $g^{\textrm{G}}$. 

Based on these preparations, we briefly explain the concept of the nonlinearity thermodynamics (NT)  without fluctuation.\cite{noltherm} In the NT, stochastic evolution of the system on configuration space, driven by the nonlinearity, is characterized by the following stochastic matrix $\mathbf{T}$:
\begin{eqnarray}
\label{eq:matT}
T_{ki} &=& R\left(q_{k}|q_{i}\right) \nonumber \\
R\left(q_{k}|q_{i}\right) &=&  z_{i}^{-1} g_{k} \exp\left[q_{k}\Gamma^{-1}q_{i} \right],
\end{eqnarray}
where $R\left(q_{k}|q_{i}\right)$ denotes transition probability from state $q_{i}$ to $q_{k}$.
Therefore $\mathbf{T}$ corresponds to the transition probability from configuration $q_{i}$ to $q_{k}$, where $g_{k}=g\left(q_{k}\right)$. From the definition of $\mathbf{T}$, we see that the matrix $\mathbf{T}$ naturally includes information about the CN at each configuration, since the $j$-th column of $T$ corresponds to the equilibrium distribution of $P_{j}$: This certainly indicates that the system transition on configuration space is driven by the nonlinearity. Based on the matrix $\mathbf{T}$, stochastic time evolution of the system is then transformed into that of thermodynamic system contacting with a thermal bath through stochastic thermodynamics, leading to e.g., deriving the nonlinearity character across multiple configurations as seen in Eq.~\eqref{eq:nth0}. 

For instance, bath entropy change through system transition from $q_{A}$ to $q_{B}$ is given by
\begin{eqnarray}
\Delta S_{\textrm{b}} = \ln \dfrac{R\left( q_{B} | q_{A} \right)}{ R\left( q_{A} | q_{B} \right) },
\end{eqnarray}
and system entropy change corresponds to its changes in terms of stochastic shanon information:
\begin{eqnarray}
\Delta S = \ln \dfrac{P\left( q_{A} \right)}{ P'\left( q_{B} \right) },
\end{eqnarray}
where $P' = \mathbf{T} P$. Hereinafter, we always employ $\Delta C$ as changes in quantity $C$ through transition from $q_{A}$ to $q_{B}$.
Through such transform with introducing (i) a new measure for the nonlinearity of 
\begin{eqnarray}
\label{eq:sen}
D_{\circ}^{K} = \sum_{q_{I}} R\left( q_{I}| q_{K} \right) \ln\dfrac{R\left( q_{I}|q_{0} \right)}{R_{\textrm{G}}\left( q_{I}|q_{0} \right) }
\end{eqnarray}
with $q_{0}$ representing a perfectly rantom configuration, and (ii) a special forward transition probability with initial states as CDOS, namely,
\begin{eqnarray}
\label{eq:p0}
P_{+}^{0} = R\left( q_{B}| q_{A} \right) g\left( q_{A} \right),
\end{eqnarray}
we obtain for instance the nonlinearity bound of Eq.~\eqref{eq:nth0}, in which the nonlinearity is measured from  $D_{\circ}$.
Now let us briefly explain the role of the introduced $D_{\circ}$ and special forward transition. From Eq.~\eqref{eq:sen}, we can clearly see that at perfectly random configuration $q_{0}$, $D_{\textrm{NOL}}^{0} - D_{\circ}^{0}=0$, which means that $D_{\circ}$ acts as the natural measure of the nonlinearity based on $q_{0}$, and for the linear system, $D_{\circ}$ always takes zero. 
For the introduced forward transition $P_{+}^{0}$, we have shown that from the basic property of the vector $H\left( q \right)$, the average $\Braket{C}_{P_{+}}^{0}$ typically corresponds to the average disparity of quantity $C$ between partially ordered and other (i.e., ordered and random) configurations for substitutional alloys.

\subsection*{Nonlinearity Thermodynamics under Fluctuation}
\begin{figure}[h]
\begin{center}
\includegraphics[width=0.98\linewidth]{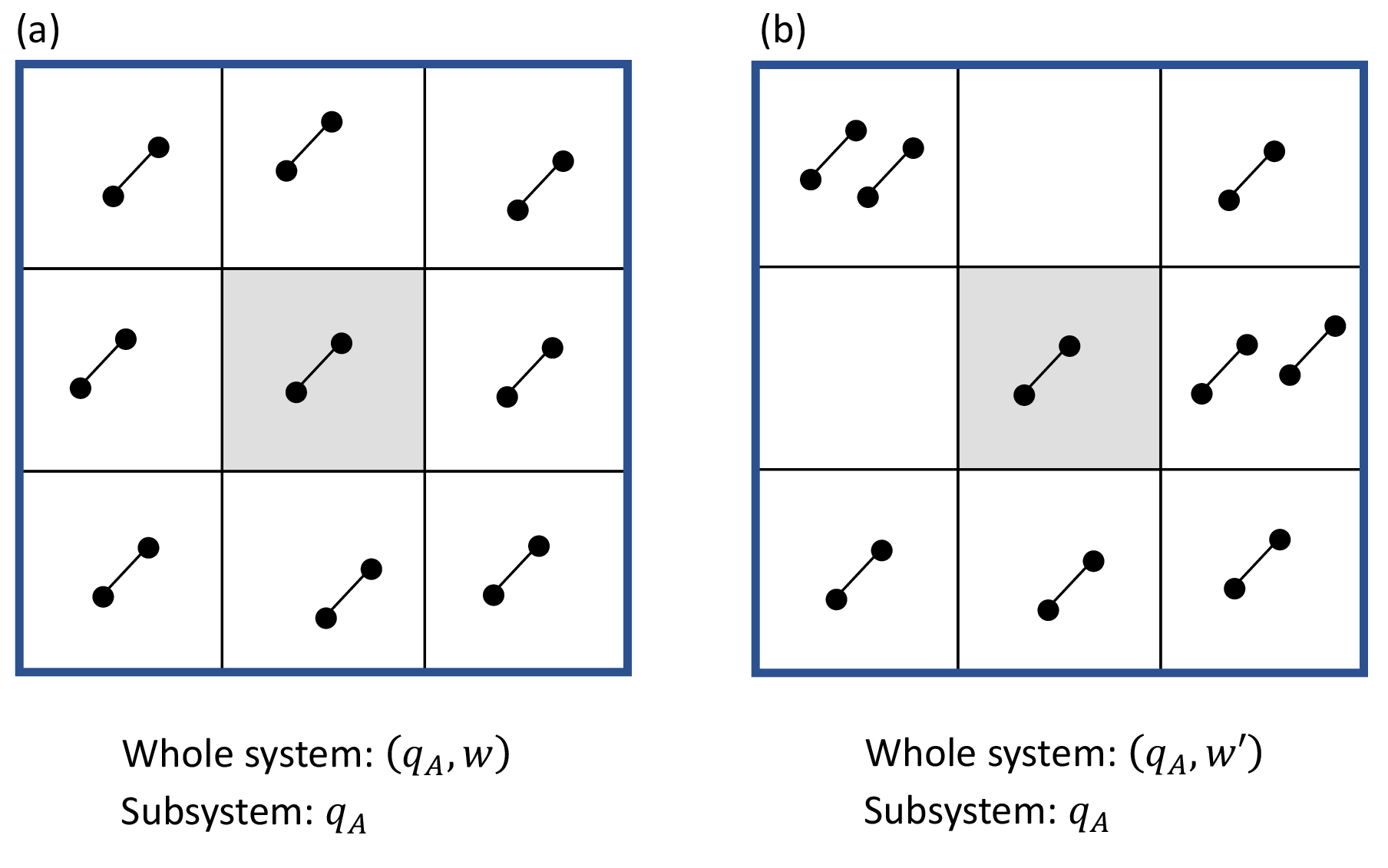}
\caption{ Schematic illustration of the correlation function for focused SDF ($\left(q_{A}\right)$) and its fluctuations ($w$ and $w'$). Whole system corresponds to that inside the bold square. }
\label{fig:ft}
\end{center}
\end{figure}
From above considerations, the present purpose is therefore to address how the thermodynamics of nonlinearity is modified when we explicitly consider the effect of ``structure fluctuation'' (SF), e.g., modification of Eq.~\eqref{eq:nth0}. To this end, we first should define the treatment of the SF in the context of the alloy configurational thermodynamics. 

Figure~\ref{fig:ft} shows the schematic iilustration of configuration for two whole systems (bold squares): Pair correlation for focused SDF takes the same value of $q_{A}$, while for r.h.s., the corresponding correlation is not uniform in subsystems, resulting in the difference in correlations for extra SDF(s) of $w\neq w'$. In Fig.~\ref{fig:ft} (b), we can see that fluctuation of $q_{A}$ appears among the subsystem. With this consideration, we here define the fluctuation $w$ for the focused SDF(s) $q_{i}$ as
\begin{eqnarray}
w = \mathbf{C}\cdot Q,
\end{eqnarray}
where $\mathbf{C}$ denotes matrix, and $Q$ is a vector. Practically, actual construction of $Q$ and $\mathbf{C}$ would depend on the  individual problems considered, which is respectively chosen as a proper set of SDFs and as a proper linear combination of the SDFs in $Q$, to capture the corresponding fluctuation in interests: For instance, one of the simplest case is that $Q$ consists of a set of long-range pair correlations, and $\mathbf{C}$ is reduced to a row vector so that the fluctuation can provide difference in configurations in Fig.~\ref{fig:ft} (a) and (b), through including a long-range correlation that is not essentially described in the subsystem. 
\begin{figure}[h]
\begin{center}
\includegraphics[width=1.00\linewidth]{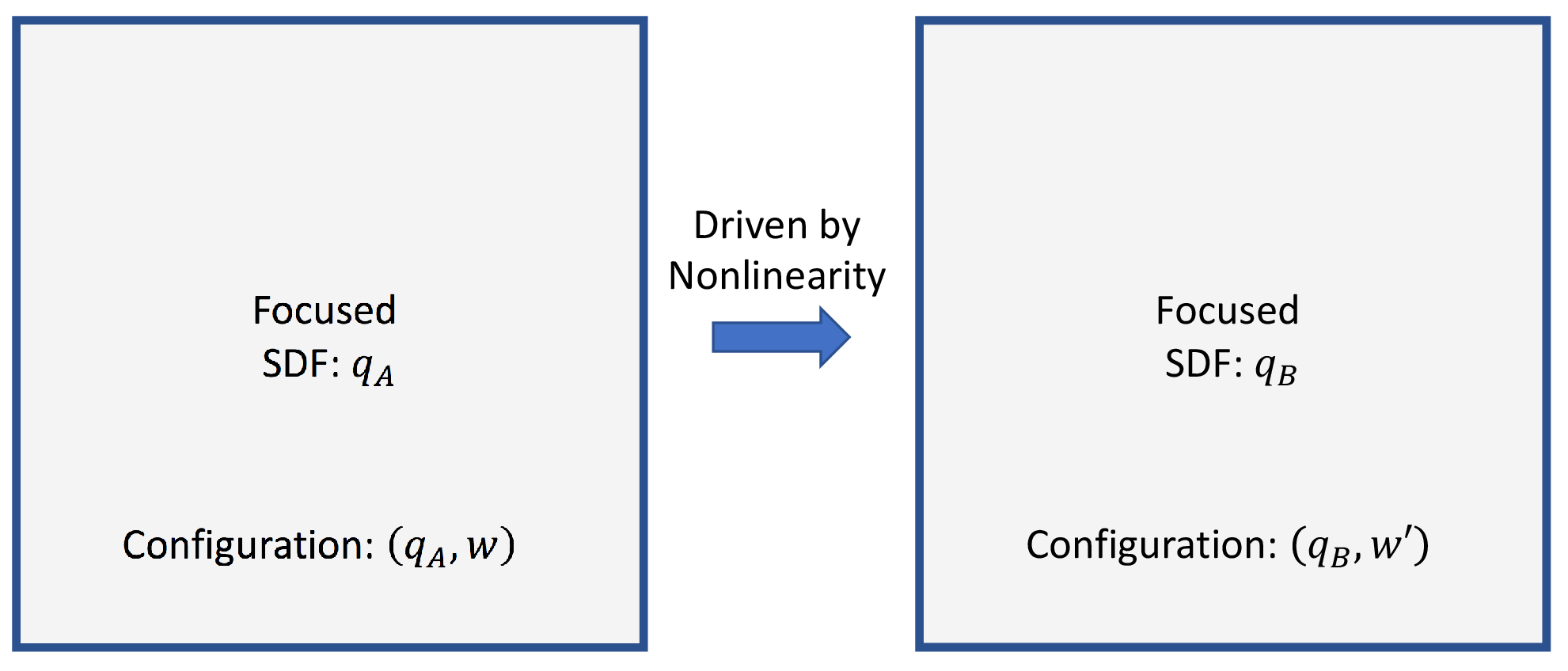}
\caption{ Schematic illustration of the evolution of whole system, driven both by the fucused SDF ($q_{A}$) and its fluctuation ($w$), resulting in the transition from $\left(q_{A}, w\right)$ to $\left(q_{B},w'\right)$.  }
\label{fig:ftnol}
\end{center}
\end{figure}

Under this setup, we now confine ourselves to the nonlinearity of the focused SDFs in Fig.~\ref{fig:ft} (a) and (b): Although the two fucused SDFs have the same correlation, $q_{A}=q_{A}$, and they also exhibit no fluctuation, their nonlinearity should essentially show different character due to the difference in fluctuation $w\neq w'$ for the whole system. We therefore set our present purpose to clarify how the nonlinearity in focused SDF is affected by the SF exsting in the whole system. Such condition holds for the practical issue, e.g.,  where the nonlinearity is described in $\left(q,w\right)$-space while physical properties in interests can be characterized within $q$-space.
In order to achieve this, we first extend the concept of nonlinearity in Eq.~\eqref{eq:asdf} to evolution of the whole system driven by the focused SDF and its fluctuation, as shown in Fig.~\ref{fig:ftnol}.

The important point here is that (i) nonlinearity of the focused SDF is described in $q$-space, while the characteristics of the nonlinearity is dominated by $\left(q,w\right)$-space. Such condition can be naturally treated through the following modified transition probability for the focused SDF under the fluctuation:
\begin{eqnarray}
\label{eq:rft}
R\left(q_{B}, w' | q_{A}, w\right) = \dfrac{g\left(q_{B}\right) \exp\left\{\left(q_{B},w'\right)\Lambda\left(q_{A},w\right)\right\} }{\sum_{q,w''}g\left(q\right) \exp\left\{\left(q,w''\right)\Lambda\left(q_{A},w\right)\right\}},
\end{eqnarray}
where $\Lambda$ denotes the inverse of $\Gamma$ for $\left(q,w\right)$-space, and we define denonimator of Eq.~\eqref{eq:rft} as  $z_{A,w} = e^{-\beta f_{A,w}}$. We can clearly see that Eq.~\eqref{eq:rft} includes modified many-body interaction $V$ in Eq.~\eqref{eq:ZV}, and the introduced $R\left(q_{B}, w' | q_{A}, w\right)$ can certainly provide different nonlinear character of the focused SDFs under different fluctuations (e.g., Fig.~\ref{fig:ft} (a) and (b)).

Starting from Eq.~\eqref{eq:rft}, we can now re-formulate the basic framework of the nonlinearity thermodynamis under the existence of fluctuation. From Eq.~\eqref{eq:rft}, the transition probability can also be interpreted as the canonical distribution under the given configuration of $q_{A}, w$, namely, $P_{A,w}\left(q_{B},w'\right)=R\left(q_{B}, w' | q_{A}, w\right)$. Therefore, the modified nonlinearity of the focused system in $q_{K}$ under $w$ is naturally given by
\begin{eqnarray}
\label{eq:nolw}
D_{\textrm{NOL}}^{K,w} = D_{\textrm{KL}}\left(P_{K,w}: P^{\textrm{G}}_{K,w}\right).
\end{eqnarray}
In a similar fashion, nonlinearity reference of $D_{\circ}$ is modified to
\begin{eqnarray}
D_{\circ}^{K,w} = \sum_{q_{i}, w''} R\left(q_{i},w''|q_{K},w\right)\ln\dfrac{R\left(q_{i},w''|q_{0},w_{0}\right)}{R_{\textrm{G}}\left(q_{i},w'|q_{0},w_{0}\right)},
\end{eqnarray}
and hereinafter the nonlinearity of Eq.~\eqref{eq:nolw} is measured from the modified $D_{\circ}$. 
We also introduce several descriptions for conveniene: $\Delta C$ always means $C$ at final configuration of $\left(q_{B}, w'\right)$ measured from $\left(q_{A},w\right)$, and $\Delta \tilde{C}$ denotes $\Delta C$ measured from that for the linear system, i.e., $\Delta \tilde{C} = \Delta C - \Delta C_{\textrm{G}}$.

Then, when the system contacts with a single thermal bath, bath entropy change can be given by
\begin{eqnarray}
\Delta S_{\textrm{b}} = \ln\dfrac{R\left(q_{B},w'|q_{A},w\right)}{R\left(q_{A},w|q_{B},w'\right)}.
\end{eqnarray}
We also introduce the followings:  
\begin{eqnarray}
\Delta d_{q} &=& d_{\textrm{KL}}\left(P_{0}\left(q_{B}\right):P_{0}^{\textrm{G}}\left(q_{B}\right)\right) - d_{\textrm{KL}}\left(P_{0}\left(q_{A}\right):P_{0}^{\textrm{G}}\left(q_{A}\right)\right) \nonumber \\
\Delta d_{w} &=& d_{\textrm{KL}}\left(P_{0}\left(w' \right):P_{0}^{\textrm{G}}\left(w'\right)\right) - d_{\textrm{KL}}\left(P_{0}\left(w\right):P_{0}^{\textrm{G}}\left(w\right)\right),  \nonumber \\
\quad
\end{eqnarray}
where $d_{\textrm{KL}}$ denotes stochastic relative entropy.
Therefore, $\Delta d_{q}$ and $\Delta d_{w}$ respectively corresponds to stochastic nonlinearity changes at the random configuration $q_{0}$ or $w_{0}$, through the transition from $\left(q_{A}, w\right)$ to $\left(q_{B},w'\right)$ for fucused SDF and fluctuation. 

Under these definitions, we can immediately obtain the following relationships: 
\begin{eqnarray}
\label{eq:dnds}
\Delta D_{\textrm{NOL}} - \Delta d_{q}  = \beta \tilde{Q},
\end{eqnarray}
where $Q$ denotes heat inflow from bath to the system. 
Eq.~\eqref{eq:dnds} certainly indicates that changes in the nonlinearity for focused SDF and heat transfer (i.e., corresponding to the bath entropy change) are connected through the additional nonlinearity information at random configuration, $\Delta d_{q}$, which is a similar characteristics for the case without fluctuation. To further address the nonlinearity character under fluctuatoin, we introduce special forward and backward transition probability for whole system where the respective initial state takes CDOS itself, namely,
\begin{eqnarray}
P_{+}^{0}\left(q_{A},q_{B},w,w'\right) &=& g\left(q_{A},w\right)R\left(q_{B},w'|q_{A},w\right) \nonumber \\
P_{-}^{0}\left(q_{A},q_{B},w,w'\right) &=& g\left(q_{B},w'\right)R\left(q_{A},w|q_{B},w'\right).
\end{eqnarray}
The above definition enables providing intuitive interpretation of its average, e.g., $\Braket{\Delta C}_{P_{+}^{0}}$ denotes average disparity of quantity $C$ between partially ordered and other (i.e. random and ordered) configurations, due to the typical characteristics of the vector $H\left(q\right)$ for substitutional alloys.  Under these transition probabilities, we can obtain the following relationship:
\begin{eqnarray}
\label{eq:sbdd}
\Delta \tilde{S}_{\textrm{b}} - \Delta d_{q} - \Delta \tilde{i} - \Delta d_{w} = \ln \dfrac{P_{+}^{0} P_{\textrm{G}-}^{0}}{P_{-}^{0}P_{\textrm{G}+}^{0}},
\end{eqnarray}
where $\Delta i$ denote difference in stochastic mutual information between focused SDF and fluctuation for the above process:
\begin{eqnarray}
\Delta i = \ln \dfrac{g\left(q_{B},w'\right)}{g\left(q_{B}\right)g\left(w'\right)} - \ln \dfrac{g\left(q_{A},w\right)}{g\left(q_{A}\right)g\left(w\right)}.
\end{eqnarray}
When we substitute Eq.~\eqref{eq:dnds} into Eq.~\eqref{eq:sbdd}, applying average of $\Braket{\quad}_{P_{+}^{0}}$ to both sides of Eq.~\eqref{eq:sbdd}, and employing Jensen's inequality for r.h.s. of Eq.~\eqref{eq:sbdd}, we finally obtain the modified bound for the nonlinearity:
\begin{eqnarray}
\Braket{\Delta D_{\textrm{NOL}}}_{P_{+}^{0}} \le \ln\Braket{ e^{-\left(\sigma_{\textrm{G}}^{0} - \Delta i_{\textrm{G}} - \Delta d_{w}\right)  }  }_{P_{+}^{0}} - \Braket{\Delta\tilde{i}+\Delta d_{w}}_{P_{+}^{0}}.
\end{eqnarray}
The above equation certainly indicates that to address the averaged nonlinearity bound in terms of the transformed thermodynamic function(s), stochastic nonlinearity information about the random configuration (i.e., $d_{w}$, $i$ and $D_{\circ}$) is somehow required. 

\section{Conclusions}
We investigate the effects of structure fluctuation on the bound for canonical nonlinearity, in the context of its thermodynamic treatment. We reveal that the upper bound for the averaged nonlinearity is characterized not only by the entropy production for linear system (fully determined by covariance matrix of the practical CDOS), but also by additional contribution due to the fluctuation, i.e., stochastic magnitude of nonseparability between focused SDF and fluctuation, and partial contribution to the nonlinearity for fluctuation at random configuration.

\section{Acknowledgement}
This work was supported by Grant-in-Aids for Scientific Research on Innovative Areas on High Entropy Alloys through the grant number JP18H05453 and  from the MEXT of Japan, and Research Grant from Hitachi Metals$\cdot$Materials Science Foundation.

\end{document}